# On the origin of high- spin states
# in nuclear fission fragments

G. Mouze , C. Ythier and S. Hachem

Faculté des Sciences, 06108 Nice cedex 2, France

In the "nucleon-phase" model of binary fission, the transfer of nucleons between an A =126 « nucleon core » and the primordial "cluster" can explain both the formation of high- spin states and the saw-tooth behavior of the variation, as a function of fragment mass, of the average angular momentum.

## 1. Introduction.

Many γ- ray properties depend on the angular momentum of the emitting system. Thus prompt γ- ray emission is very useful for studying the initial value of fission-fragment spins. In such a study, Armbruster et al. [1] made the interesting observation that the variation, as a function of fragment mass, of the average value, $J_{r.m.s.}$, of the angular momentum created in the neutron-induced fission of $^{235}U$ has a saw-tooth behavior. In fact, a similar saw-tooth behavior had been previously observed in the prompt- neutron emission of fission fragments as a function of their mass [2,3] and had been carefully analyzed by J. Terrell [4], who proposed the following relation for the mean number $\bar{\nu}$ of emitted neutrons as a function of fragment mass:

$$\bar{\nu} = 0.08 (A_L - 82) + 0.10 (A_H - 126), \tag{1}$$

a relation holding for the n-induced fission of $^{233}U$, $^{235}U$ and $^{239}Pu$, and the spontaneous fission of $^{252}Cf$.

However, De Frenne [5] reports that the $J_{r.m.s.}$- values obtained e.g. by Denschlag et al.[6] and Bocquet et al. [7] from measured isomeric ratios in the n-induced fission of $^{235}U$ do not support the idea of a saw-tooth behavior.

Fortunately, great advances have now been made in the spectroscopy of neutron-rich fission fragments thanks to the advent of γ- ray arrays using a great number of germanium detectors, such as Gamma-sphere [8,9]. And thanks to elaborate methods of high-fold γ- ray coincidences, detailed and accurate decay schemes can be established today. Moreover, spins and parities of the excited states can now be carefully determined. Thus it is possible to measure reliably the average angular momentum.

Since the nucleon -phase model of binary fission is based on Terrell's discovery of 1962 [4], it may be asked whether this model can lead to a better understanding of the pioneer work of Armbruster et al. [1] and explain the origin of the angular moment of fission fragments. It is the aim of the present communication.

## 2. Durell's study of the average J of fission fragments produced in $^{235}U + n_{th}$.

In a careful prompt- γ-ray study of these fission fragments -- a study in which corrections have been made for the angular - momentum reduction caused by neutron emission and by statistical γ- ray emission -- J.L. Durell made in 1996 the important observation that the $J_{r.m.s.}$-value increases as a function of fragment mass, both for light and heavy fragments [10]. Fig.1 reports his results, presented at the 3$^{rd}$ Conference on Dynamical Aspects of Nuclear Fission. The straight lines in this figure are our own interpretation.

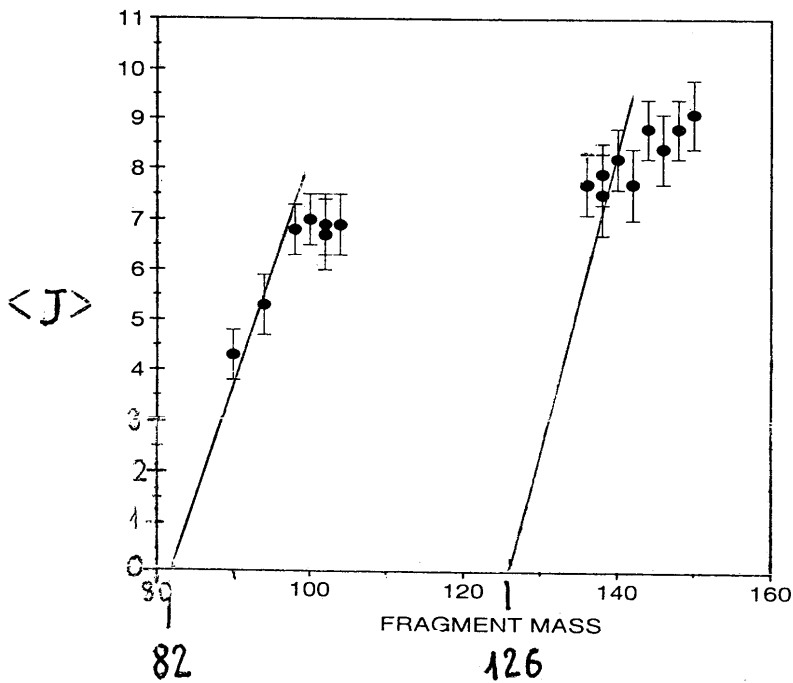

Fig.1 Variation, as a function of fragment mass, of the angular momentum of fission fragments measured by J.L. Durell [10] in the neutron-induced fission of $^{235}U$. This figure is adapted from fig.4 of ref.[10] with permission. The straight lines are our own interpretation.

It is easy to justify that the straight line joining the $J_{r.m.s}$ values of the fragments of mass A = 94,96 and 98 is well represented by equation (2)

$$J_{r.m.s}(\hbar) = 0.458\,(A_L - 82). \qquad (2)$$

The slope of the red straight line corresponding to the heavy fragments is more arbitrary, because data are lacking; indeed, high-spin isomers prevent an accurate study of prompt transitions in this region. But one sees that a straight line joining $J_{r.m.s}= 0$, at $A =126$, to the J- value of fragments of mass $A =140$ has a slope about 0.573 ℏ, and that *these slopes, 0.458 ℏ and 0.573 ℏ, are in the same ratio as those of Terrell's relation, namely 0.8.*

Interestingly, the slope of the light-fragment branch does not differ much from that of the work of Armbruster et al., namely ~ 0.518 ℏ.

These observations are an encouragement to search an explanation in the frame of the nucleon-phase model.

## 3. The nucleon-phase model.

At the 2008 Winter Meeting on Nuclear Physics [11], we suggested an explanation of Terrell's law, based on the hypothesis of the formation of hard "nucleon cores", made of 82 and 126 nucleons, in the rearrangement step of binary fission. A review of these ideas has been given in ref.[12].

In this model, binary fission results from an initial collision within the fissioning system between the primordial $^{208}$Pb core and a cluster, of mass $A_{cl}$, made of its valence nucleons (e.g. a $^{28}$Ne nucleus, in the n-induced fission of $^{235}$U) [13]: this collision creates an "A = 126 nucleon core" in $^{208}$Pb and releases 82 "free nucleons", which can be transferred to the cluster. But this cluster has a great tendency to form with them an "A = 82 nucleon core". After the capture, by this cluster, of a number (82 – $A_{cl}$) of free nucleons, the remaining 82 – (82 – $A_{cl}$) = $A_{cl}$ nucleons are shared between the free states of the valence shells of the A = 82 nucleon core and the free states of the valence shells of the A = 126 nucleon core, according to a statistical distribution law which has some resemblance to the law describing the sharing-out of one and the same body between two unmiscible solvents [14].

But what happens to the first nucleons, in excess of the (82 – $A_{cl}$) nucleons captured by the primordial cluster, as they are projected, with a great moment $\vec{p}$, in the direction of the transfer axis, towards the new hard A = 82 nucleon core?

They can create a great "angular momentum" $\vec{r} \wedge \vec{p}$ when they hit the nascent fragment of mass A > 82, at a distance $\vec{r}$ of its center of mass.

Similarly, the free nucleons of the nucleon phase can create a great angular momentum when they hit the nascent heavy fragment, of mass A > 126, at a distance $\vec{r}$ of its center of mass, with a great moment $\vec{p}$. At the beginning, the valence shells of the new A = 82 and A = 126 nucleon cores are empty;

consequently, the angular momentum is expected to increase from the value zero, at A =82 and A = 126, up to higher values at higher masses of the fragments; and this increase is expected to be linear, as a function of the fragment mass, at low values of the mass; but, due to various effects, the $J_{r.m.s}$ - value is expected to increase more slowly at higher fragment masses.

This situation might be compared to that of the excitation of high angular momenta in the capture of heavy ions by a nucleus, realized for the first time in the famous Morinaga-Gugelot experiment [15].

An interpretation of the relations $J_{r.m.s}$= 0.458 ($A_L$ − 82) $\hbar$ and $J_{r.m.s}$= ~ 0.573 ($A_H$ - 126) $\hbar$ is that the created angular momentum is proportional to the mass of the fragments in excess of A = 82 and A = 126, i.e. proportional to the mass of the "soft" valence shells of the nascent fragments. Similarly, Terrell's relations

$\bar{\nu}_L$ = 0.08 ($A_L$ − 82) and $\bar{\nu}_H$ = 0.10 ($A_H$ − 126), mean that the number of emitted neutrons is proportional to the number of nucleons present in the "open" valence shells of the nascent fragments.

## 4. Conclusion

The variation, as a function of mass, of the average angular momentum of fission fragments, now furnished by prompt γ–ray studies such as that of Durell, really follows a law similar to Terrell's law of 1962, a law which involves "magic mass numbers", 82 and 126 [11]. We have shown that the fragment angular momentum can result from the capture, by the "valence shells" of two hard nucleon cores, made of 82 and 126 nucleons, of part of the [ 82 − (82 − $A_{cl}$)] = $A_{cl}$ free nucleons, and that the angular momentum increases linearly with the mass of these valence shells at the beginning of their filling. This proportionality to the mass of the soft valence shells is quite understandable for angular momenta.